# Comparison of Deep Learning Techniques on Human Activity Recognition using Ankle Inertial Signals

Farhad Nazari*, *Member*, IEEE, Darius Nahavandi, *Member*, IEEE, Navid Mohajer, and Abbas Khosravi, *Senior Member,* IEEE

*Institute for Intelligent System Research and Innovation (IISRI), Deakin University, Australia*

*Abstract— Human Activity Recognition (HAR) is one of the fundamental building blocks of human assistive devices like orthoses and exoskeletons. There are different approaches to HAR depending on the application. Numerous studies have been focused on improving them by optimising input data or classification algorithms. However, most of these studies have been focused on applications like security and monitoring, smart devices, the internet of things, etc. On the other hand, HAR can help adjust and control wearable assistive devices, yet there has not been enough research facilitating its implementation. In this study, we propose several models to predict four activities from inertial sensors located in the ankle area of a lower-leg assistive device user. This choice is because they do not need to be attached to the user's skin and can be directly implemented inside the control unit of the device. The proposed models are based on Artificial Neural Networks and could achieve up to 92.8% average classification accuracy.*

## I. Introduction

Human Activity Recognition (HAR) is one of the fundamental building blocks of many human-in-the-centre technologies like orthoses and exoskeletons. A wearable assistive device [1], in its simplest form, needs to adjust itself with the user's activity and posture constantly. On the other hand, an elaborate system requires working around its user's complex ranges of motion by predicting the intended action or movement of the user [2]. In both cases, accurately detecting the user's activity can help with the control or adjustment of the device by either switching between predefined and tuned settings or activating sub-level control systems.

There are different approaches to HAR depending on the application and available input information. While image processing methods are being used for detecting unusual activity recognition [3], signal processing methods are widely adopted in applications that use time-series signals like physical activity detection using smartphones [4] and watches [5] sensors. The applications extend to healthcare [6], remote monitoring [7], smart homes [8], human-robot interaction [9], [10], gaming [11] and other data mining applications [12]. Some researchers have recently tried to adopt these techniques for wearable assistive devices. Poliero et al. used HAR to enhance the back-support exoskeleton's versatility [13]. Zheng et al. implemented a Decision Tree (DT) classifier on surface EMG sensors to detect the human torso's motion [14].

There are various algorithms and mathematical methods to map the input data to the desired class. In the past two decades, supervised Machine Learning (ML) techniques like DT [15], K-Nearest Neighbors (KNN) [16], Gradient Boosting (GB) [17], Linear Discriminant Analysis (LDA) [18] and Support Vector Machine (SVM) [19] have been widely used in human activity recognition. The downside of these traditional ML-based approaches is their high reliance on feature extraction techniques without a systematic or universal approach [20]. With the increase in computing power through GPUs and parallel computing in recent years, deep learning approaches are being adopted in the field. They are capable of extracting high-level features from raw or processed signals. Oniga and Suto conducted a series of studies to optimise the sensor configuration and recognition systems' complexity and found that a two-layer feed-forward back-propagation Artificial Neural Network (ANN) with a sigmoid activation function yields good results and is easy to implement on FPGA [21]. Inoue et al. showed that a Deep Recurrent Neural Network (DRNN) could achieve up to 71% better recognition rate compared to traditional ML models [22].

The increased computation power allowed researchers to combine different strategies to improve performance. Cho and Yoon proposed a hierarchical one-dimensional Convolutional Neural Network (1D CNN) with data sharpening to enhance HAR performance [23]. In another attempt, Xia et al. achieved up to 95.8% $F_1$ score in their LSTM-CNN model by stacking convolution layers after two LSTM [24]. Dua et al. demonstrated up to 97.2% classification accuracy by combining CNN and Gated Recurrent Unit (GRU) [25].

Despite the large number of research conducted in HAR, there aren't nearly enough studies related to the activity detection for the application of wearable assistive devices. Hence, most developed methodologies and knowledge may not be directly transferrable to this field. This limitation can stem from the type of information, sensor location, feature extraction detection approaches, and predefined activities. In this study, we aim to detect the human activity for the application of a lower-limb assistive device, e.g. foot/ankle prosthesis/ exoskeleton. The proposed models have been trained on the Inertial Measurement Unit (IMU) data located at the ankle. The advantage of this approach is that the sensor does not require to be attached to the user's skin. Moreover, it can be implemented in the control unit of the device, increasing the speed and response rate and reducing the complexity. The activities are among frequently mundane

* corresponding author: f.nazari@deakin.edu.au

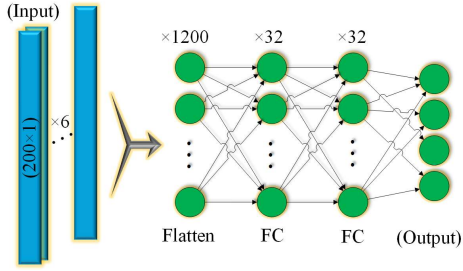

*Fig. 1. The structure of the DL model.*

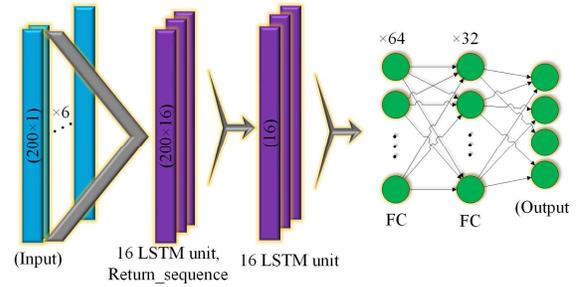

*Fig. 2. The structure of the LSTM model.*

activities that require different functionality from the assistive device.

The next section discusses our methodology and proposed models. In section III, the dataset and data processing methods will be described. Then the results will be discussed in section VI, followed by the conclusion in the last section.

## II. MODEL DEVELOPMENT

In this study, we propose various Deep Learning (DL) models to detect human activity from the inertial data related to the ankle motion of an assistive device. A shallow learning model has also been developed as the baseline model for benchmarking. These models have been trained on a random two-second portion of ankle motion in a k-fold cross-validation method, in which k is the number of subjects. The train and test split are based on the participant to prevent data leakage. Adam optimiser and early stopping methods have been used to achieve the best results. Categorical crossentropy has been chosen as the loss function and a high dropout rate of 0.3 is added to each layer to prevent overfitting. All other hyper-parameters have been optimised by trial and error. The proposed models are as follows:

### 1) Baseline Shallow NN model:

This is the simplest model, comprising an input layer (200×6 nodes) without any activation function and an output layer (four classes) with the activation function being a normalised exponential function or SoftMax. This model has 4,804 trainable parameters.

### 2) Deep Neural Networks (DL) model:

This model adds two Fully Connected (FC) layers with 32 neurons and Rectified Linear Unit (ReLU) activation function to the shallow NN model, allowing to pick deeper and more complex patterns in the data. This network has 39,620 trainable parameters. Fig 1 shows the structure of the DL model.

### 3) Recurrent Neural Networks (Simple RNN) model:

RNNs are a class of networks that allows previous outputs to be used as inputs while having a hidden state. This is useful when the input is a sequence like time-series. The other benefit of RNN layer is that it can get a sequence as an input and return a single output, reducing the overall size of the network. The proposed RNN model consists of two simple-RNN layers followed by two FC-layers with 64 and 32 neurons respectively, followed by a four-neuron output layer, leading to total 7,652 trainable parameters. All layers except the output use ReLU activation function.

### 4) Long Short-Term Memory (LSTM) model:

Simple RNN networks suffer from the vanishing gradient problem. LSTM solves this issue by introducing another hidden state for long-term memory. The proposed LSTM model is similar to the simple RNN one, with the different unit sizes in the first two layers (16 instead of 32) leading to 6,884 trainable parameters. All layers use the hyperbolic tangent (tanh) activation function except the input and output layers. Fig 2 shows the structure of the proposed LSTM model.

### 5) Gated Recurrent Units (GRU) model:

GRU is another modification to the RNN hidden layer to help with capturing a more extended range of connections in the time-series data. It combines the long- and short-term memory states into one state. The architecture of the GRU model is similar to the simple RNN one, with the only difference being the tanh activation function. This network has 14,500 trainable parameters.

### 6) Convolutional Neural Network (CNN) model:

CNNs are a class of deep learning networks that work by detecting the local features of the input data. Some examples of these features in time-series signals are increase or decrease in the value, slope change, zero crossing, etc. The network is capable of extracting higher-level features by stacking up multiple convolutional layers. Fig 3 shows the structure of the proposed CNN model. This network uses the ReLU activation function in all layers except the output layer and has 51,308 trainable parameters.

### 7) Combined convolution-recurrent NN model:

By stacking convolutional and recurrent neural layers, the network can use the properties of both networks. Here, we combined the first two layers of the proposed CNN model with three RNN ones. These models use the ReLU activation function for the convolutional, softmax for output, and tanh for the other layers. CNN-SimpleRNN, CNN-GRU, and CNN-LSTM have 14,836, 23,348 and 27,316 trainable parameters.

## III. DATASET AND DATA PROCESSING

This section describes the dataset selection and processing methods used in this study.

### A. Dataset and Data Selection

Some portion of the PAMAP2 Physical Activity Monitoring dataset has been used in this study [26]. In this dataset, nine people performed 18 different activities while wearing three IMU sensors on the wrist, chest and ankle and a heart rate monitor. IMU signals have been recorded with 100Hz sampling rate and contain temperature, 3D acceleration, gyroscope and magnetometer data.

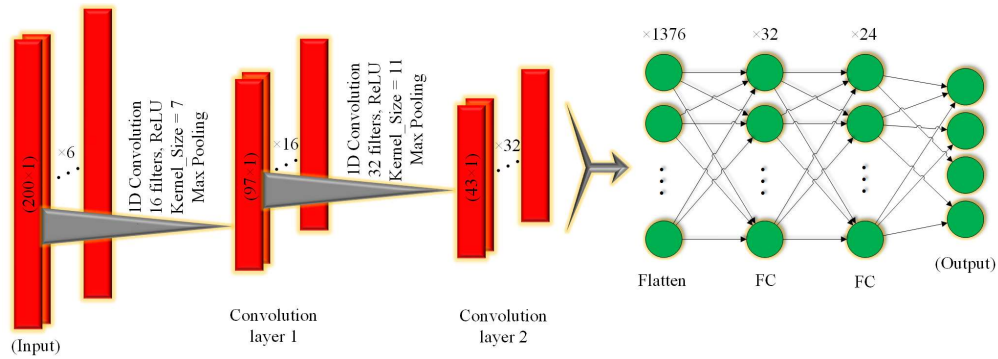

Fig. 3. The structure of the proposed CNN model.

The advantage of this dataset is that it contains the sensor data related to a set of activities relevant to the lower limb prosthesis/ exoskeleton. We only used the accelerometer and gyrometer data of the dominant ankle movements for this study. The temperature doesn't seem to vary in a meaningful way during activities. Also, due to the need for regular calibration of the magnetometer, we decided that this type of sensor might not be the best candidate for our application. Location-wise, having sensors on the chest and wrist adds some complications. Each sensor needs a separate processor, power supply and communication protocol, which is usually wireless, reducing the sampling frequency and increasing the latency. Not only do they increase the set up complexity and costs, they also add more points of failure. What is more, they can have different latencies due to inherit lags in the reading, transmission and processing of each sensor's signals. We ignored the sensor data related to the chest and wrist for these reasons. Needless to say that in multi-sensor set-ups wrong location set up of the sensors by the user can harm the performance, whereas in our proposed model the sensor can be fixed in the device.

*B. Data Preparation*

In the experiment set up by Reiss and Stricker [27], nine subjects' participation in 18 different categories is not uniform. Also, not all activities have a meaningful difference in the assistive device's functionality. Considering the above, we decided to choose four activities from eight participants. These activities are standing, walking, stair ascending and descending.

After mapping the signals to mean zero and standard deviation of one, the multi-modal time-series data has been transformed into 3D features for deep learning algorithms with a window size of two seconds. The reason for that is that we wanted the signals to be long enough to have meaningful information related to human activity while being short enough to be able to update it in real-time. Each window has a 60ms slide, meaning that it has a 140ms overlap with the previous window.

## IV. RESULTS AND DISCUSSION

All the proposed models have been trained in an 8-fold cross-validation method, in which train and test split has been done based on one of the eight selected subjects. Table 1 shows the average 8-fold cross-validation accuracy of the proposed models. The baseline model (Shallow_NN) performed with 84.6% average accuracy. Adding two fully-connected layers improves this number for the DL model by 4.6%. This is because the extra layers allow for the detection of deeper and more complex patterns. While adding simple RNN layers hurt the performance of the DL model, GRU and LSTM layers improved the performance[1] by 3.5 and 8.4%, respectively. The vanishing gradient problem could be the main culprit for performance drop in the simple RNN model. These layers cannot retain old information over long time-series data. While GRU solves this problem by retaining both short and long-term memory, LSTM allows for better detail capturing by having them in separate hidden states. However, the CNN model showed the best performance among the proposed models, with an average accuracy of 92.8%. This is due to the superb ability of convolutional layers to detect local features. Fig 4 shows the distribution of accuracy results of the proposed models over eight training rounds.

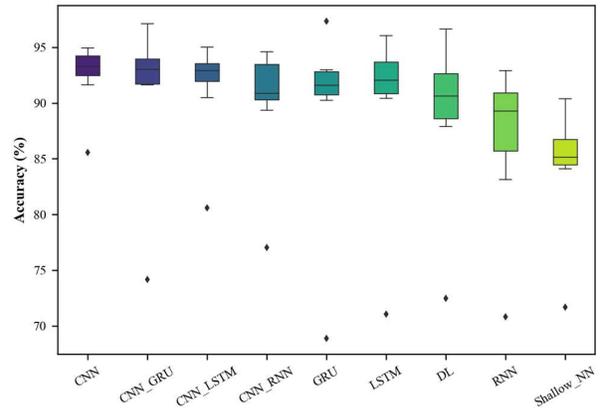

Fig. 4. Accuracy of the proposed models over eight training rounds.

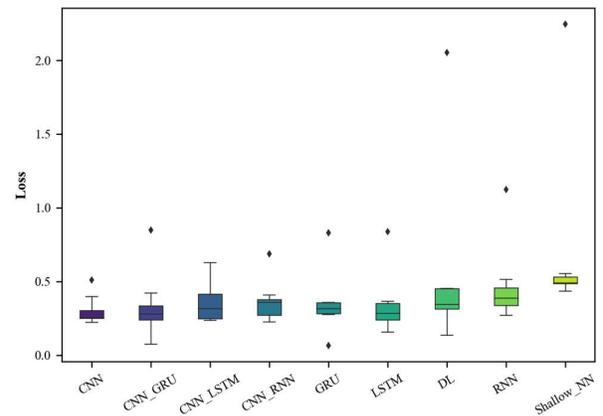

Fig. 5. Loss value of of the proposed models over eight training rounds. The loss function is crossentropy.

*TABLE 1. The average 8-fold cross-validation loss and accuracy of the proposed models.*

| Proposed Model | Average Accuracy (%) | CrossEntropy Loss |
|---|---|---|
| CNN | 92.8 | 0.30 |
| CNN_LSTM | 91.5 | 0.36 |
| CNN_GRU | 91.1 | 0.34 |
| CNN_RNN | 90.1 | 0.37 |
| LSTM | 90.1 | 0.34 |
| GRU | 89.5 | 0.35 |
| DL | 89.2 | 0.55 |
| RNN | 86.9 | 0.47 |
| Shallow_NN | 84.6 | 0.71 |

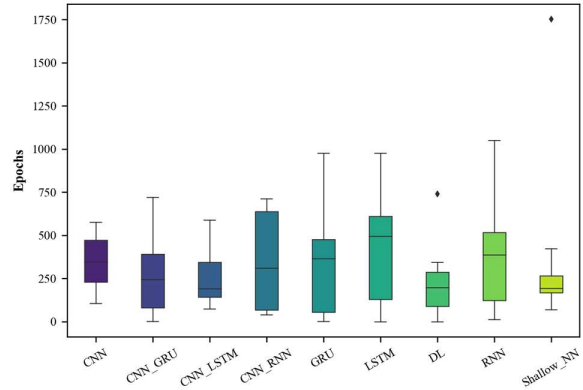

*Fig. 6. Number of epochs required to train the proposed models.*

In terms of consistency, CNN showed the narrowest range of accuracies with a 3% standard deviation and a 9.4% difference between the best and worst results. These numbers are slightly higher for the CNN_LSTM model at 4.6 and 14.4%, respectively. This is surprising as the LSTM model showed around 73% higher deviation in accuracy results than the CNN_LSTM model. The least consistent model was GRU which was almost three times less consistent than the CNN model.

The CNN model showed the lowest cross-entropy loss values averaging at 0.3, followed by CNN_GRU and LSTM by 11 and 14% higher. GRU and other combinational models are in the same range. The Shallow_NN model has the highest loss, followed by the DL model. Fig 5 shows the distribution of cross-entropy loss values of the proposed models over eight training rounds.

In terms of training, the baseline model, on average, took 393.5 rounds to train. This number was highest for the LSTM model at 425, followed by the simple_RNN model at 400.5. CNN and GRU models needed an average of 347 training epochs to train. DL and CNN_LSTM models were the fastest models in training time by an average of 241.5 and 256.8 epochs. Fig 6 shows the distribution of the number of epochs required to train the proposed models.

It was predictable that feeding extracted local features to fully connected layers could increase the performance. We hypothesised that adding convolutional layers before recurrent ones will boost the accuracy by feeding the extracted features instead of raw signals to the network. The results approve this hypothesis for all RNN networks with 24.7, 14.8 and 14.4% improvement in the performance of Simple_RNN, GRU and LSTM, respectively. However, the performance drop in combinational networks compared to the CNN model was not anticipated. It looks like allowing the deep neural networks to detect higher levels of features from the local ones extracted by CNN can yield better results than further processing this sequence of features in RNNs.

## V. CONCLUSION

In this study, we proposed and evaluated various artificial neural networks and models to detect human activity from an inertial sensor, embedded in a lower leg assistive device. Four activities have been chosen to have a meaningful effect on the adjustment or control of the device that is slow-paced and present in daily activities: standing, walking, ascending and descending stairs. The proposed models are a baseline shallow neural networks, multi-layer perceptron or deep neural networks, simple recurrent neural networks, gated recurrent units, long short-term memory, convolutional neural network and combinational models.

The models were trained on two-second portions of accelerometer and gyrometer data and evaluated with the k-fold cross-validation method, while the train and test split was based on participants to prevent data leakage. The convolutional model demonstrated the best performance, averaging 92.8% accuracy, followed by combinational models. Shallow and simple recurrent neural network models had the lowest average accuracies at 84.6 and 86.9%, respectively. We were expecting to observe better results from combinational models. However, they showed slightly inferior performance compared to the convolutional model. This drop in performance can be investigated in future works.

For future work, the sensor type and location [28] can be optimised for better performance, while the potential for detailed phase recognition can be studied for the target activity [29]. Human body position [30], EEG methodologies [31] and emotion recognition [32] can be used as complementary to the aforementioned models. Furthermore, these models could be used to activate the proper control system for robotic devices [33]–[37]. Also, these devices provide the user with haptic feedback [38]. In addition to that, autonomous systems [39], control [40] and learning [41] and virtual reality simulations [42] could be further investigated.

---

[1] Decrease in the wrong guesses